\documentclass[twocolumn,showpacs,preprintnumbers,nofootinbib,prd,superscriptaddress,groupedaddress,10pt]{revtex4-1}
\usepackage{graphicx,amssymb,amsmath,amsthm,amsfonts,epsfig,epsf}
\usepackage{hyperref}
\usepackage{cleveref}
\usepackage[usenames]{color}
\usepackage{epstopdf}

\usepackage{booktabs,array}
\usepackage[british]{babel}

\usepackage{aas_macros}
\usepackage{bm}
\usepackage{dcolumn}
\usepackage[latin1]{inputenc}
\usepackage{latexsym}
\usepackage{rotating}
\usepackage{color}
\usepackage{longtable}
\usepackage{enumerate}
\usepackage{tensor}
\usepackage{url}
\def\f{\frac}

\def\Flm{{\cal F}_{lmn}}
\def\flm{f_{lmn}}

\newcommand{\ben}{\begin{enumerate}}
\newcommand{\een}{\end{enumerate}}

\def\be{\begin{equation}}
\def\ee{\end{equation}}
\def\bea{\begin{eqnarray}}
\def\eea{\end{eqnarray}}

\newcommand{\beq}{\begin{eqnarray}}
\newcommand{\eeq}{\end{eqnarray}} 
\newcommand{\ba}{\begin{align}}
\newcommand{\ea}{\end{align}}

\begin{document}

\title{Spectroscopy of Kerr black holes with Earth- and space-based interferometers}

\author{
Emanuele Berti$^{1,2}$,
Alberto Sesana$^{3}$,
Enrico Barausse$^{4,5}$,
Vitor Cardoso$^{2,6}$, 
Krzysztof Belczynski$^{7}$ 
}

\affiliation{${^1}$ Department of Physics and Astronomy, The University of 
Mississippi, University, MS 38677, USA}
\affiliation{${^2}$ CENTRA, Departamento de F\'isica, Instituto Superior
T\'ecnico, Universidade de Lisboa, Avenida Rovisco Pais 1,
1049 Lisboa, Portugal}
\affiliation{${^3}$ School of Physics and Astronomy, The University of Birmingham, 
Edgbaston, Birmingham B15 2TT, UK}
\affiliation{${^4}$ Sorbonne Universit\'es, UPMC Universit\'e Paris 6, UMR 7095, 
Institut d'Astrophysique de Paris, 98 bis Bd Arago, 75014 Paris, France}
\affiliation{${^5}$ CNRS, UMR 7095, Institut d'Astrophysique de Paris, 98 bis Bd Arago, 
75014 Paris, France}
\affiliation{${^6}$ Perimeter Institute for Theoretical Physics, 31 Caroline Street North
Waterloo, Ontario N2L 2Y5, Canada}
\affiliation{${^7}$ Astronomical Observatory, Warsaw University,
  Al. Ujazdowskie 4, 00-478 Warsaw, Poland}

\begin{abstract}
  We estimate the potential of present and future interferometric
  gravitational-wave detectors to test the Kerr nature of black holes
  through ``gravitational spectroscopy,'' i.e. the measurement of
  multiple quasinormal mode frequencies from the remnant of a black
  hole merger. Using population synthesis models of the formation and
  evolution of stellar-mass black hole binaries, we find that
  Voyager-class interferometers will be necessary to perform these
  tests. Gravitational spectroscopy in the local Universe may become
  routine with the Einstein Telescope, but a 40-km facility like
  Cosmic Explorer is necessary to go beyond $z\sim 3$.  In contrast,
  eLISA-like detectors should carry out a few -- or even hundreds --
  of these tests every year, depending on uncertainties in massive
  black hole formation models. Many space-based spectroscopical
  measurements will occur at high redshift, testing the strong gravity
  dynamics of Kerr black holes in domains where cosmological
  corrections to general relativity (if they occur in nature) must be
  significant.
\end{abstract}

%\tableofcontents
%\end{widetext}
%\clearpage

%\pacs{04.70.-s,04.25.dc}
%95.35.+d 	Dark matter (stellar, interstellar, galactic, and cosmological) (see also 95.30.Cq Elementary particle processes; for brown dwarfs, see 97.20.Vs; for galactic halos, see 98.35.Gi or 98.62.Gq; for models of the early %Universe, see 98.80.Cq)
%14.80.-j 	Other particles (including hypothetical)
%11.10.St 	Bound and unstable states; Bethe-Salpeter equations
%12.60.-i 	Models beyond the standard model (for unified field theories, see 12.10.-g)
%04.25.D-    Numerical relativity
%04.25.dc    Numerical studies of critical behavior, singularities, and cosmic censorship
%04.25.dg    Numerical studies of black holes and black-hole binaries
%04.25.-g    general relativity: approximation methods, equations of motion
%04.40.-b 	Self-gravitating systems; continuous media and classical fields in curved spacetime
%04.50.-h    Higher-dimensional gravity and other theories of gravity
%04.50.Cd    Kaluza-Klein theories
%04.50.Gh    Higher-dimensional black holes, black strings, and related objects
%04.60.Cf    Gravitational aspects of string theory
%04.70.-s    Physics of black holes
%04.70.Bw    Classical black holes
%04.70.Dy    Quantum aspects of black holes, evaporation, thermodynamics
%04.80.-y    Experimental studies of gravity
%04.80.Cc    Experimental tests of gravitational theories
%11.25.Mj    Compactification and four-dimensional models
%11.10.Kk    Field theories in dimensions other than four

\maketitle

%%%%%%%%%%%%%%%%%%%%%%%%%%%%%%%%%%%%%%%%%%%%%%%%%%%%%%%%%%%%%%%%%%%%%%%%%%%
\noindent{\bf{\em Introduction.}}
%%%%%%%%%%%%%%%%%%%%%%%%%%%%%%%%%%%%%%%%%%%%%%%%%%%%%%%%%%%%%%%%%%%%%%%%%%%
%
The first binary black hole (BH) merger signal detected by the LIGO
Scientific Collaboration, GW150914~\cite{Abbott:2016blz}, had a
surprisingly high combined signal-to-noise ratio (SNR) of $24$ in the
Hanford and Livingston detectors. The quasinormal mode signal
(``ringdown'') from the merger remnant is consistent with the
predictions of general relativity (GR) for a Kerr BH, but it was
observed with a relatively low SNR
$\rho\sim 7$~\cite{TheLIGOScientific:2016src}.  The large masses of
the binary components~\cite{TheLIGOScientific:2016wfe} have
interesting implications for the astrophysics of binary BH
formation~\cite{TheLIGOScientific:2016htt}. This detection, together
with a second detected BH merger~\cite{Abbott:2016nmj}, placed
interesting constraints on the merger rates of BH binaries in the Universe
\cite{Dominik:2014yma,Belczynski:2015tba,Abbott:2016nhf,Belczynski:2016obo,TheLIGOScientific:2016pea}.

LISA Pathfinder was successfully launched in December 2015, paving the
way for a space-based detector such as
eLISA~\cite{AmaroSeoane:2012km,AmaroSeoane:2012je}, which will observe
mergers of massive BHs throughout the Universe with very large SNRs
and test the Kerr nature of the merger remnants. The basic idea is
that the dominant $\ell=m=2$ resonant frequency and damping time can
be used to determine the remnant's mass $M$ and dimensionless spin
$j=J/M^2$ (we adopt geometrical units $G=c=1$ throughout this Letter.)
In GR, all subdominant mode frequencies (e.g. the modes with
$\ell=m=3$ and $\ell=m=4$~\cite{Berti:2007fi}) are then uniquely
determined by $M$ and $j$. The detection of subdominant modes requires
high SNR, but each mode will provide one (or more) tests of the Kerr
nature of the remnant~\cite{Berti:2009kk}.  As first pointed out by
Detweiler in 1980, gravitational waves allow us to do BH spectroscopy:
``After the advent of gravitational wave astronomy, the observation of
these resonant frequencies might finally provide direct evidence of
BHs with the same certainty as, say, the 21 cm line identifies
interstellar hydrogen''~\cite{Detweiler:1980gk}.

Such high SNRs are known to be achievable with an eLISA-like
detector~\cite{Berti:2005ys}.  The surprisingly high SNR of GW150914
raised the question whether current detectors at design sensitivity
should routinely observe ringdown signals loud enough to perform
gravitational spectroscopy. Leaving aside conceptual issues about
ruling out exotic
alternatives~\cite{Damour:2007ap,Barausse:2014tra,Cardoso:2016rao},
here we use our current best understanding of the astrophysics of
stellar-mass and supermassive BHs to compute the rates of events that
would allow us to carry out spectroscopical tests.

\begin{figure*} \centering
  \includegraphics[width=0.8\textwidth,clip=true,angle=0]{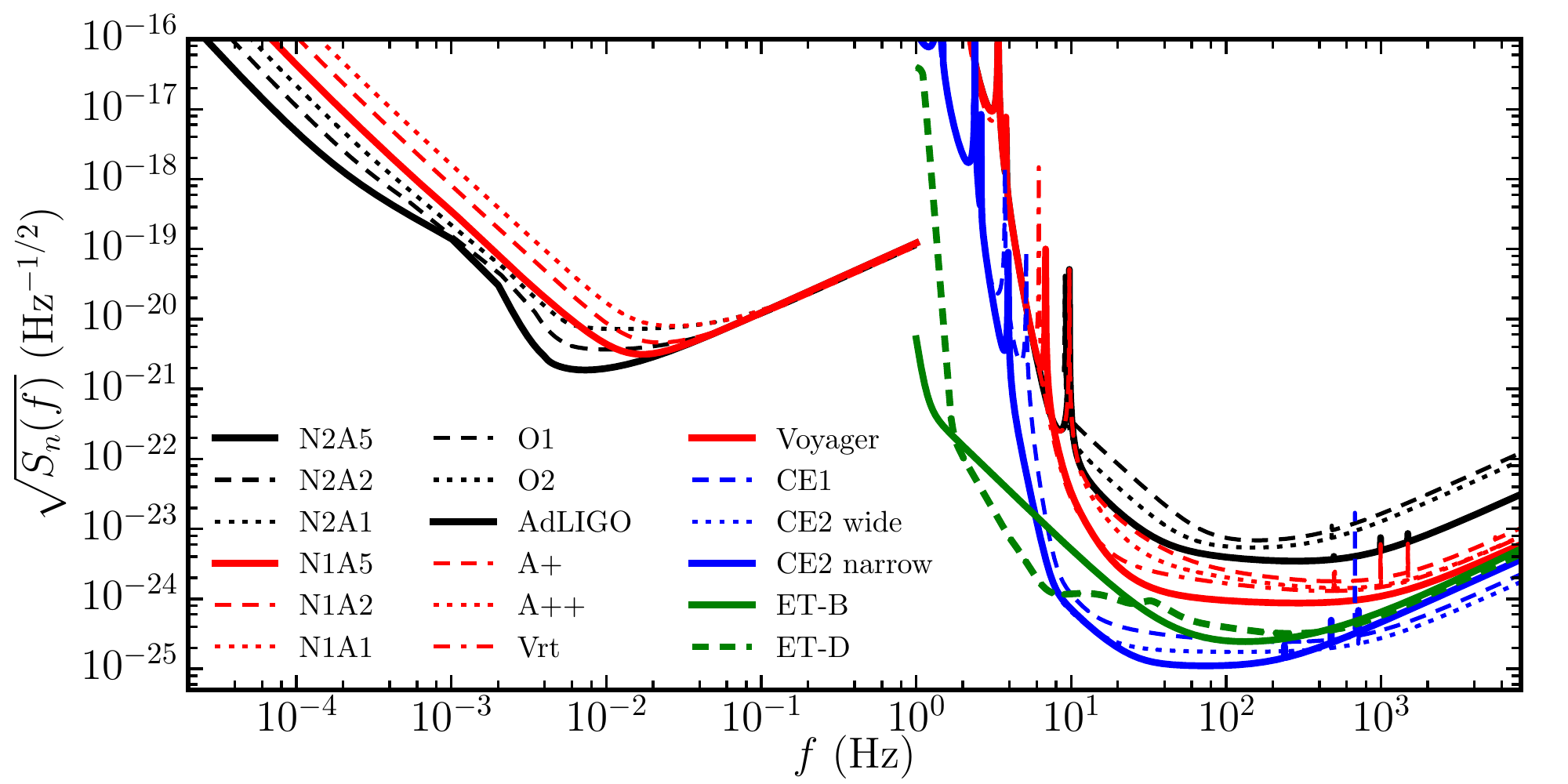}
  \caption{\label{fig:thirdgen} Noise PSDs for various space-based and
    advanced Earth-based detector designs. ``N$i$A$k$'' refers to non
    sky-averaged eLISA PSDs with pessimistic (N1) and optimistic (N2)
    acceleration noise and armlength $L=k$~Gm
    (cf.~\cite{Klein:2015hvg}). In the high-frequency regime, we show
    noise PSDs for (top to bottom): the first AdLIGO observing run
    (O1); the expected sensitivity for the second observing run (O2)
    and the Advanced LIGO design sensitivity
    (AdLIGO)~\cite{Aasi:2013wya}; the pessimistic and optimistic
    ranges of AdLIGO designs with squeezing (A+,
    A++)~\cite{Miller:2014kma} ; Vrt and
    Voyager~\cite{Adhikari:2013kya,Voyager}; Cosmic Explorer (CE1),
    basically A+ in a 40-km facility~\cite{Dwyer:2015bma}; CE2 wide
    and CE2 narrow, i.e. 40-km detectors with Voyager-type technology
    but different signal extraction tuning~\cite{Dwyer:PC,Voyager};
    and two possible Einstein Telescope designs, namely
    ET-B~\cite{ETweb} and ET-D in the ``xylophone''
    configuration~\cite{Hild:2009ns}.}
\end{figure*}

Below we provide the details of our analysis, but the main conclusions
can be understood relying on the noise power spectral densities (PSDs)
$S_n(f)$ of present and future detectors, as shown and briefly
reviewed in Fig.~\ref{fig:thirdgen}, and simple back-of-the-envelope
estimates.

%%%%%%%%%%%%%%%%%%%%%%%%%%
%%%%%%%%%%%%%%%%%%%%%%%%%%

%%%%%%%%%%%%%%%%%%%%%%%%%%%%%%%%%%%%%%%%%%%%%%%%%%%%%%%%%%%%%%%%%%%%%%%%%%%
\noindent{\bf{\em Ringdown SNR.}}
%%%%%%%%%%%%%%%%%%%%%%%%%%%%%%%%%%%%%%%%%%%%%%%%%%%%%%%%%%%%%%%%%%%%%%%%%%%
Consider the merger of two BHs with source-frame masses $(m_1,\,m_2)$,
spins $({\bf j}_1,\,{\bf j}_2)$, total mass $M_{\rm tot}=m_1+m_2$,
mass ratio $q\equiv m_1/m_2\geq 1$ and symmetric mass ratio
$\eta=m_1m_2/M_{\rm tot}^2$.  The remnant mass and dimensionless spin,
$M$ and $j=J/M^2$, can be computed using the fitting formulas in
\cite{Barausse:2012qz} and \cite{Hofmann:2016yih}, respectively (see
also \cite{Rezzolla:2007rz,Barausse:2009uz}).  The ringdown SNR $\rho$
can be estimated by following~\cite{Berti:2005ys}.  Including redshift
factors and substituting the Euclidean distance $r$ by the luminosity
distance $D_L$ as appropriate, Eq.~(3.16) of~\cite{Berti:2005ys}
implies that $\rho$ is well approximated by
\bea
\label{rhoanalytic}
\rho=\f{\delta_{\rm eq}}{D_{\rm L}\Flm}\left[
  \f{8}{5}\f{M_z^3\epsilon_{\rm rd}}{S_n(\flm)} \right]^{1/2}\,,
\eea
where $M_z=M(1+z)$.  Fits of the mass-independent dimensionless
frequencies $\Flm(j) \equiv 2\pi M_z\flm$
%and quality factor $\Qlm(j)$
are given in Eq.~(E1) of~\cite{Berti:2005ys}. 
The geometrical factor $\delta_{\rm eq}=1$ for Michelson
interferometers with orthogonal arms. For eLISA-like detectors the
angle between the arms is 60$^\circ$, so $\delta_{\rm eq}=\sqrt{3}/2$,
and we use the {\it non sky-averaged} noise PSD
$S_n(f)$~\cite{Berti:2004bd,Klein:2015hvg}.
%, and we have used the approximation $4\Qlm\gg 1$.
%
The ringdown efficiency for nonspinning binaries is well approximated
by the matched-filtering estimate of Eq.~(4.17) in
\cite{Berti:2007fi}: $\epsilon_{\rm rd}=0.44\eta^2$.  When using the
best-fit parameters inferred for
GW150914~\cite{TheLIGOScientific:2016wfe}, Eq.~(\ref{rhoanalytic})
yields a ringdown SNR $\rho\simeq 7.7$ in O1 (in agreement
with~\cite{TheLIGOScientific:2016src}) and $\rho\simeq 16.2$ in
AdLIGO.

Due to the orbital hang-up effect~\cite{Campanelli:2006uy}, spinning
binaries with aligned (antialigned) spins radiate more (less) than
their nonspinning counterparts. The dominant spin-induced correction
to the radiated energy is proportional to a weighted sum of the components of
the binary spins along the orbital angular
momentum~\cite{Boyle:2007sz,Boyle:2007ru,Barausse:2012qz}. We estimate
this correction by rescaling the radiated energy by the factor
$E_{\rm rad}(m_1,\,m_2,\,{\bf j}_1,\,{\bf j}_2)/E_{\rm
  rad}(m_1,\,m_2,\,{\bf 0},\,{\bf 0})$,
where the total energy radiated in the merger $E_{\rm rad}$ is
computed using Eq.~(18) of~\cite{Barausse:2012qz}. We find that
spin-dependent corrections change $\rho$ by at most 50\%.

It is now easy to understand why Einstein Telescope-class detectors
are needed to match the SNR of eLISA-like detectors and to perform BH
spectroscopy.  The quantity $\Flm(j)$ is a number of order
unity~\cite{Berti:2005ys,Berti:2009kk}. The physical frequency is
$\flm\propto 1/M_z$: for example, an equal-mass merger of nonspinning
BHs produces a remnant with $j\simeq 0.6864$ and fundamental ringdown
frequency $f_{220} \simeq 170.2 (10^2\,M_\odot/M_z)$~Hz.  So
Earth-based detectors are most sensitive to the ringdown of BHs with
$M_z\sim 10^2M_\odot$, while space-based detectors are most sensitive
to the ringdown of BHs with $M_z\sim 10^6M_\odot$. The crucial point
is that, according to Eq.~\eqref{rhoanalytic}, $\rho\sim M^{3/2}$ at
fixed redshift and noise PSD. As shown in Fig.~\ref{fig:thirdgen}, the
``bucket'' of the N2A5 eLISA detector is at
$S_{\rm N2A5}^{1/2}\sim 10^{-21}$~Hz$^{-1/2}$. This noise level is
$\sim 10^2$ ($10^3$, $10^4$) times larger than the best sensitivity of
AdLIGO (Voyager, Einstein Telescope), respectively.  However eLISA BHs
are $\sim 10^4$ times more massive, yielding signal amplitudes that
are larger by a factor $\sim 10^6$. Astrophysical rate calculations
are very different in the two frequency regimes, but these qualitative
arguments explain why only Einstein Telescope-class detectors will
achieve SNRs nearly comparable to eLISA.

%%%%%%%%%%%%%%%%%%%%%%%%%%
\begin{figure*}[t] 
\centering
  \includegraphics[width=0.95\columnwidth,clip=true,angle=0]{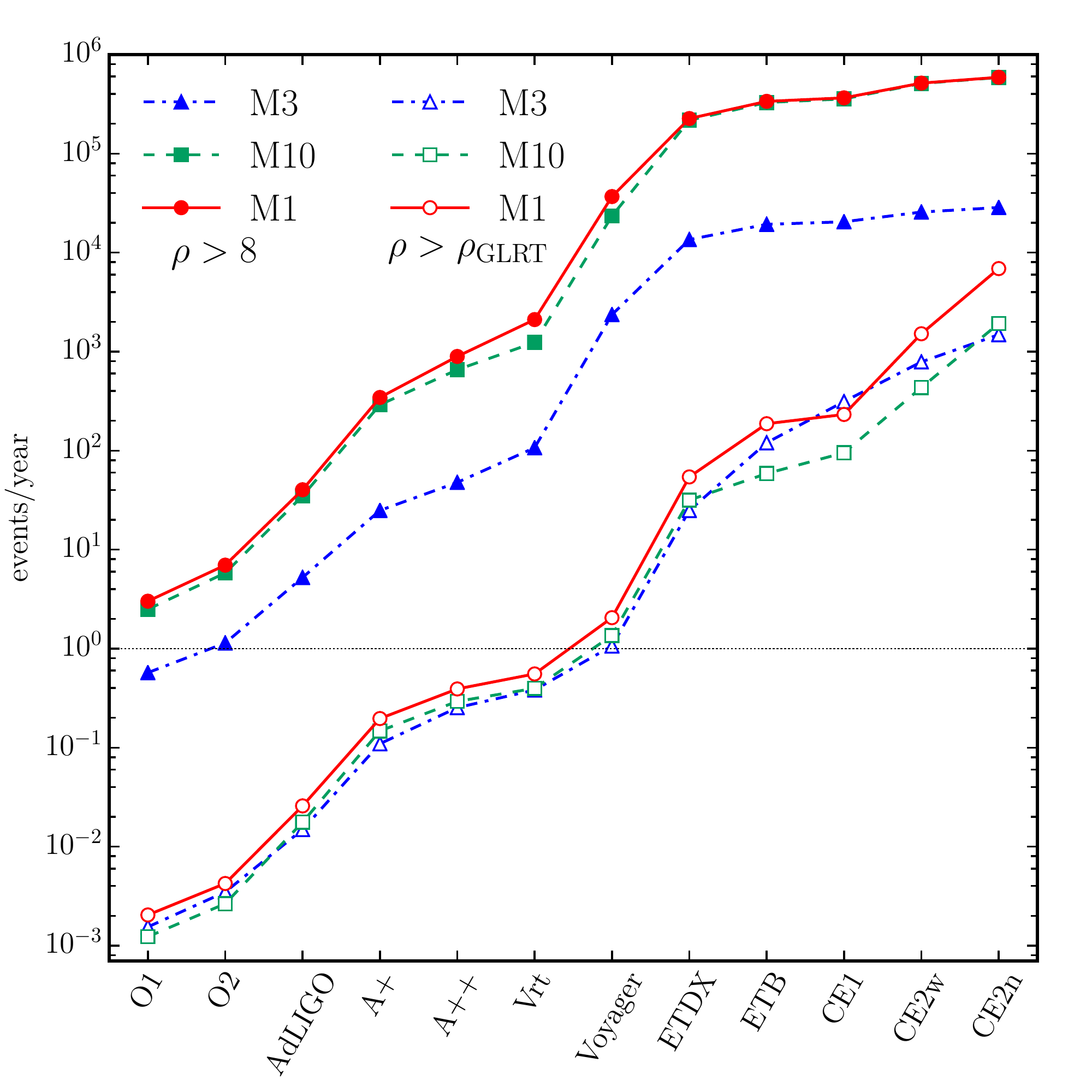}
  \includegraphics[width=0.95\columnwidth,clip=true,angle=0]{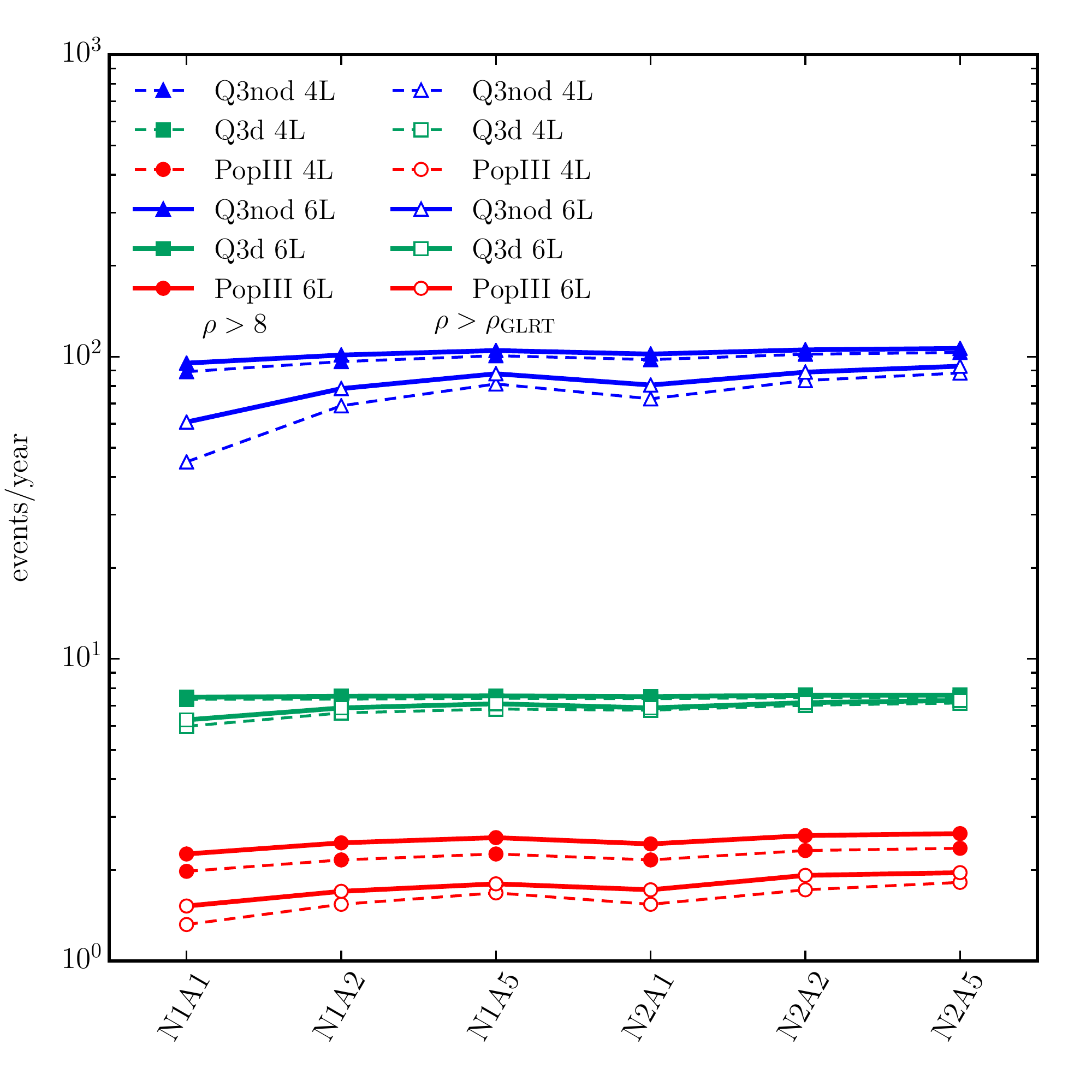}
  \caption{\label{fig:ratesGLRT} Rates of binary BH mergers that yield
    detectable ringdown signals (filled symbols) and allow for
    spectroscopical tests (hollow symbols). Left panel: rates per year
    for Earth-based detectors of increasing sensitivity. Right panel:
    rates per year for 6-link (solid) and 4-link (dashed) eLISA
    configurations with varying armlength and acceleration noise.}
\end{figure*}
%%%%%%%%%%%%%%%%%%%%%%%%%%

%%%%%%%%%%%%%%%%%%%%%%%%%%%%%%%%%%%%%%%%%%%%%%%%%%%%%%%%%%%%%%%%%%%%%%%%%%%
\noindent{\bf{\em Astrophysical models.}}
%%%%%%%%%%%%%%%%%%%%%%%%%%%%%%%%%%%%%%%%%%%%%%%%%%%%%%%%%%%%%%%%%%%%%%%%%%%
We estimate {\em ringdown} detection rates for Earth-based
interferometers (detection rates for the full inspiral-merger-ringdown
signal are higher) using three population synthesis models computed
with the {\tt Startrack} code: models M1, M3 and M10.  Models M1 and
M3 are the ``standard'' and ``pessimistic'' models described
in~\cite{Belczynski:2016obo}.
The ``standard model'' M1 and model M10 predict very similar rates for
AdLIGO at design sensitivity.  In both of these models, compact
objects receive natal kicks that decrease with the compact object
mass, with the most massive BHs receiving no natal kicks.  This
decreases the probability of massive BHs being ejected from the
binary, increasing merger rates. Model M1 allows for BH masses as high
as $\sim 100~M_\odot$. On the contrary, model M10 includes the effect
of pair-instability mass loss, which sets an upper limit of
$\sim 50M_\odot$ on the mass of stellar origin
BHs~\cite{Belczynski:2016jno}.
In model M3, all compact objects (including BHs) experience high natal
kicks drawn from a Maxwellian with $\sigma=265$km\, s$^{-1}$ based on
the natal kick distribution measured for single pulsars in our
Galaxy~\cite{Hobbs:2005yx}.  The assumption of large natal kicks leads
to a severe reduction of BH-BH merger rates, and therefore model M3
should be regarded as pessimistic~\cite{Belczynski:2016obo}.
In all of these models we set the BH spins to zero, an assumption
consistent with estimates from
GW150914~\cite{TheLIGOScientific:2016htt}. Even in the unrealistic
scenario where all BHs in the Universe were maximally spinning, rates
would increase by a factor $\lesssim 3$~(see Table 2 of
\cite{Dominik:2014yma}). Massive binaries with ringdowns detectable by
Earth-based interferometers could also be produced by other mechanisms
(see
e.g.~\cite{Benacquista:2011kv,Rodriguez:2016kxx,Marchant:2016wow,deMink:2016vkw}),
and therefore our rates should be seen as lower bounds.

To estimate ringdown rates from massive BH mergers detectable by eLISA
we consider the same three models (PopIII, Q3nod and Q3d) used
in~\cite{Klein:2015hvg} and produced with the semi-analytical approach
of \cite{Barausse:2012fy} (with incremental improvements described in
\cite{Sesana:2014bea,letter,newpaper}). These models were chosen to
span the major sources of uncertainty affecting eLISA rates, namely
(i) the nature of primordial BH seeds (light seeds coming from the
collapse of Pop III stars in model PopIII; heavy seeds originating
from protogalactic disks in models Q3d and Q3nod), and (ii) the delay
between galaxy mergers and the merger of the BHs at galactic centers
(model Q3d includes this delay; model Q3nod does not, and therefore
yields higher detection rates). In all three models the BH spin
evolution is followed
self-consistently~\cite{Barausse:2012fy,Sesana:2014bea}. For each
event in the catalog we compute $\rho$ from Eq.~(\ref{rhoanalytic}),
where $\epsilon_{\rm rd}$ is rescaled by a spin-dependent factor as
necessary.

%%%%%%%%%%%%%%%%%%%%%%%%%%%%%%%%%%%%%%%%%%%%%%%%%%%%%%%%%%%%%%%%%%%%%%%%%%%
\noindent{\bf{\em Detection rates.}}
%%%%%%%%%%%%%%%%%%%%%%%%%%%%%%%%%%%%%%%%%%%%%%%%%%%%%%%%%%%%%%%%%%%%%%%%%%%
%
The ringdown detection rates (events per year with $\rho>8$ in a
single detector) predicted by models M1, M3, M10 (for stellar-mass BH
binaries) and PopIII, Q3d, Q3nod (for supermassive BH binaries) are
shown in Fig.~\ref{fig:ratesGLRT} with filled symbols. For example,
models M1 (M10, M3) predict $3.0$ ($2.5$, $0.57$) events per year with
detectable ringdown in O1; 7.0 (5.8, 1.1) in O2; and 40 (35, 5.2) in
AdLIGO. Model Q3d (Q3nod, PopIII) predicts 38 (533, 13) events for a
6-link N2A5 eLISA mission lasting 5 years, but in the plot we divided
these numbers by 5 to facilitate a more fair comparison in terms of
events {\it per year}.

%%%%%%%%%%%%%%%%%%%%%%%%%%
\begin{figure*}[t] \centering
  \includegraphics[width=0.95\columnwidth,clip=true,angle=0]{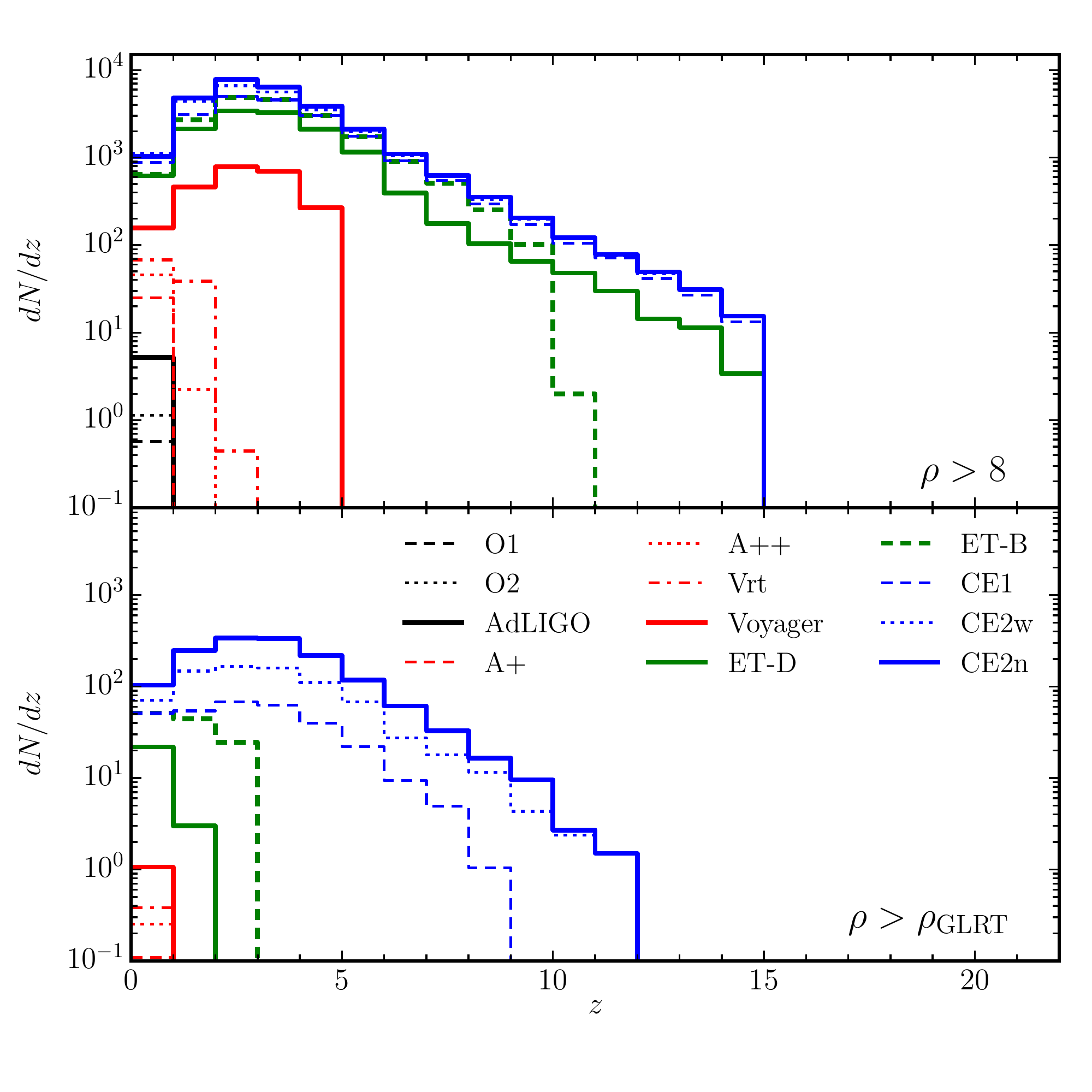}
  \includegraphics[width=0.95\columnwidth,clip=true,angle=0]{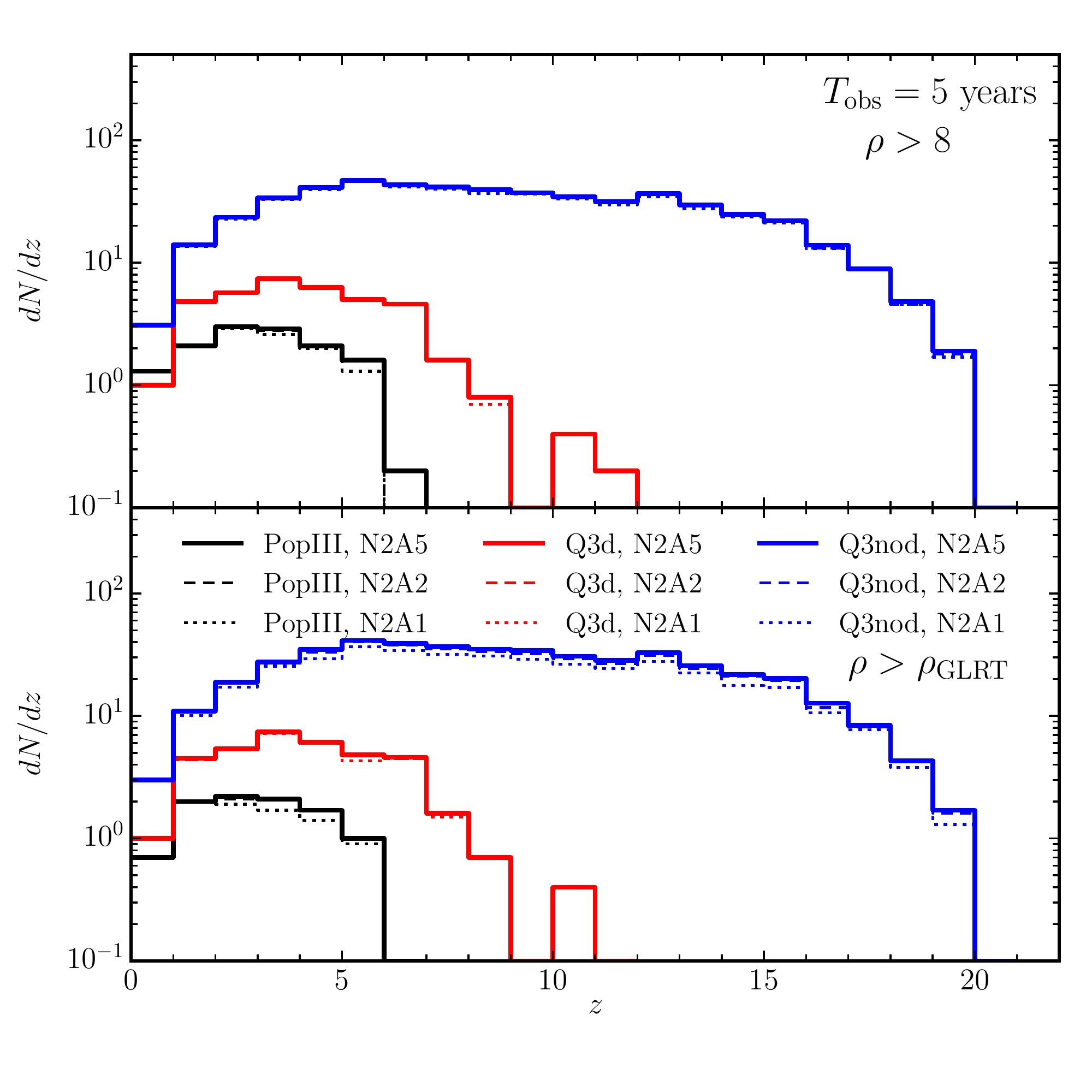}
  \caption{\label{fig:redshift_eLISA_Earth} Left: redshift
    distribution of events with $\rho>8$ (top) and
    $\rho>\rho_{\rm GLRT}$ (bottom) for model M1 and Earth-based
    detectors. In the bottom-left panel, the estimated AdLIGO rate
    ($\approx 2.6\times 10^{-2}$~events/year) is too low to display.
    Right: same for models Q3nod, Q3d and PopIII.  Different eLISA
    design choices have an almost irrelevant impact on the
    distributions.}
\end{figure*}
%%%%%%%%%%%%%%%%%%%%%%%%%%

%%%%%%%%%%%%%%%%%%%%%%%%%%%%%%%%%%%%%%%%%%%%%%%%%%%%%%%%%%%%%%%%%%%%%%%%%%%
\noindent{\bf{\em BH spectroscopy.}}
%%%%%%%%%%%%%%%%%%%%%%%%%%%%%%%%%%%%%%%%%%%%%%%%%%%%%%%%%%%%%%%%%%%%%%%%%%%
Suppose that we know that a signal contains two (or possibly more)
ringdown modes.  We expect the weaker mode to be hard to resolve if
its amplitude is low and/or if the detector's noise is large. The
critical SNR for the second mode to be resolvable can be computed
using the generalized likelihood ratio test (GLRT)~\cite{Berti:2007zu}
under the following assumptions: (i) using other criteria, we have
already decided in favor of the presence of one ringdown signal; (ii)
the ringdown frequencies and damping times, as well as the amplitude
of the dominant mode, are known. Then the critical SNR
$\rho_{\rm GLRT}$ to resolve a mode with either $\ell=m=3$ or
$\ell=m=4$ from the dominant mode with $\ell=m=2$ is well fitted, for
nonspinning binary BH mergers, by
\beq
\rho_{\rm GLRT}^{2,\,3}&=&17.687+\frac{15.4597}{q-1} -\frac{1.65242}{q}\,,\\
\rho_{\rm GLRT}^{2,\,4}&=&37.9181+\frac{83.5778}{q} +\frac{44.1125}{q^2} + \frac{50.1316}{q^3}\,.
\eeq
These fits reproduce the numerical results in Fig.~9 of
\cite{Berti:2007zu} within $0.3\%$ when $q \in [1.01-100]$.
Spectroscopical tests of the Kerr metric can be performed whenever
either mode is resolvable, i.e.
$\rho>\rho_{\rm GLRT}\equiv \min(\rho_{\rm GLRT}^{2,\,3}, \rho_{\rm
  GLRT}^{2,\,4})$.
The $\ell=m=3$ mode is usually easier to resolve than the $\ell=m=4$
mode, but the situation is reversed in the comparable-mass limit
$q\to 1$, where the amplitude of odd-$m$ modes is
suppressed~\cite{Berti:2007fi,London:2014cma}.
Extreme mass-ratio calculations~\cite{Barausse:2011kb} and a
preliminary analysis of numerical waveforms show that the ratio of
mode amplitudes is, to a good accuracy, spin-independent, therefore
this SNR threshold is adequate for our present purpose. 

The rates of events with $\rho>\rho_{\rm GLRT}$ are shown in
Fig.~\ref{fig:ratesGLRT} by curves with hollow symbols. The key
observation here is that, although ringdown {\it detections} should be
routine already in AdLIGO, high-SNR events are exceedingly rare:
reaching the threshold of $\sim 1$ event/year requires Voyager-class
detectors, while sensitivities comparable to Einstein Telescope are
needed to carry out such tests routinely. This is not the case for
space-based interferometers: typical ringdown detections have such
high SNR that $\approx 50\%$ or more of them can be used to do BH
spectroscopy.
The total number of eLISA detections and spectroscopic tests depends
on the underlying BH formation model, but it is remarkably independent
of detector design (although the N1A1 design would sensibly reduce
rates in the most optimistic models).

Perhaps the most striking difference between Earth- and space-based
detectors is that a very large fraction of the ``spectroscopically
significant'' events will occur at cosmological redshift in eLISA, but
not in Einstein telescope. This is shown very clearly in
Fig.~\ref{fig:redshift_eLISA_Earth}, where we plot redshift histograms
of detected events (top panel) and of events that allow for
spectroscopy (bottom panel). eLISA can do spectroscopy out to
$z\approx 5$ (10, or even 20!) for PopIII (Q3d, Q3nod) models, while
even the Einstein Telescope is limited to $z\lesssim 3$. Only 40-km
detectors with cosmological reach, such as Cosmic
Explorer~\cite{Dwyer:2015bma,Dwyer:PC}, would be able to do
spectroscopy at $z\approx 10$.

%%%%%%%%%%%%%%%%%%%%%%%%%%%%%%%%%%%%%%%%%%%%%%%%%%%%%%%%%%%%%%%%%%%%%%%%%%%
\noindent{\bf{\em Conclusions.}}
%%%%%%%%%%%%%%%%%%%%%%%%%%%%%%%%%%%%%%%%%%%%%%%%%%%%%%%%%%%%%%%%%%%%%%%%%%%
%
Using our best understanding of the formation of field binaries, we
predict that AdLIGO at design sensitivity should observe several
ringdown events per year. However routine spectroscopical tests of the
dynamics of Kerr BHs will require the construction and operation of
detectors such as the Einstein
Telescope~\cite{Sathyaprakash:2012jk,Gossan:2011ha,Meidam:2014jpa},
and 40-km detectors~\cite{Dwyer:2015bma,Dwyer:PC} will be necessary to
reach cosmological distances.
Many of the mergers for which eLISA can do BH spectroscopy will be
located at $z\gg 1$. These systems will test GR in qualitatively
different regimes than any low-$z$ observation by AdLIGO: BH
spectroscopy with eLISA will test whether gravity behaves {\it
  locally} like GR even at the very early epochs of our Universe,
possibly placing constraints on proposed extensions of Einstein's
theory~\cite{Gair:2012nm,Yunes:2013dva,Berti:2015itd,Yunes:2016jcc}.

Given the time lines for the construction and operation of these
detectors, it is likely that the first instances of BH spectroscopy
will come from a space-based detector.  This conclusion is based on
the simple GLRT criterion introduced in~\cite{Berti:2007zu}, and it is
possible that better data analysis techniques (such as the Bayesian
methods advocated in~\cite{Gossan:2011ha,Meidam:2014jpa}) could
improve our prospects for gravitational spectroscopy with Earth-based
interferometers. We hope that our work will stimulate the development
of these techniques and their use on actual data.

As shown in Fig.~\ref{fig:ratesGLRT}, differences in rates between
models M1 and M10 become large enough to be detectable in A+. We
estimate 34 (29) ringdown events per year for M1 (M10) in A+, and 89
(66) events per year in A++. Rate differences are even larger when we
consider the complete signal. Therefore, while the implementation of
squeezing in AdLIGO may not allow for routine BH spectroscopy, it
could reveal the nature of the BH mass spectrum in the range
$\sim [50-100]~M_\odot$.

%%%%%%%%%%%%%%%%%%%%%%%%%%%%%%%%%%%%%%%%%%%%%%%%%%%%%%%%%%%%%%%%%%%%%%%%%%%%%%
\noindent{\bf{\em Acknowledgments.}}
%%%%%%%%%%%%%%%%%%%%%%%%%%%%%%%%%%%%%%%%%%%%%%%%%%%%%%%%%%%%%%%%%%%%%%%%%%%%%%
%\begin{acknowledgments}
E.~Berti thanks Scott Hughes and Stefano Vitale for exchanges that
motivated this work, and Sheila Dwyer for sharing noise PSD data for
advanced detectors.
E.~Berti was supported by NSF CAREER Grant No.  PHY-1055103, by NSF
Grant No. PHY-1607130 and by FCT contract IF/00797/2014/CP1214/CT0012
under the IF2014 Programme.
A.S. is supported by a University Research Fellowship of the Royal
Society.
E. Barausse acknowledges support from the European Union's Seventh
Framework Programme (FP7/PEOPLE-2011-CIG) through the Marie Curie
Career Integration Grant GALFORMBHS PCIG11-GA- 2012-321608.
V.C. acknowledges financial support provided under the European
Union's H2020 ERC Consolidator Grant ``Matter and strong-field
gravity: New frontiers in Einstein's theory'' grant agreement
no. MaGRaTh--646597.
K.B.  acknowledges support from the NCN grant Sonata Bis 2
(DEC-2012/07/E/ST9/01360).
Research at Perimeter Institute is supported by the Government of
Canada through Industry Canada and by the Province of Ontario through
the Ministry of Economic Development $\&$ Innovation.
This work was supported by the H2020-MSCA-RISE-2015 Grant
No. StronGrHEP-690904.
We would like to thank thousands of Universe@home users that have
provided their personal computers for our simulations.

%\bibliographystyle{apsrev4-1}
%\bibliography{eLISARD}


\end{document}